\documentclass[usegraphicx,usedcolumn,usenatbib,letterpaper]{mn2e}

\usepackage{color,epsfig,amssymb,longtable}
\usepackage{array,colortbl,lscape}
\newcommand{\apj}{ApJ}

\bibpunct[,]{(}{)}{;}{a}{,}{,}

\begin{document}

\title[Star formation scenarios for star-forming dwarf galaxies]{Evolution of star forming dwarf galaxies: Characterizing the star formation scenarios}

\author[{Mart\'{\i}n-Manj\'{o}n} et al.]  {M.L. Mart\'{\i}n-Manj\'{o}n$^{1}$ \thanks{E-mail:mariluz.martin@uam.es}, 
M.~Moll\'{a},$^{2}$, A.~I.~D\'{\i}az$^{1,3}$, and R.~Terlevich$^{3}$ \thanks{Research Affiliate 
IoA, Cambridge} \\ $^{1}$Departamento
de F\'{\i}sica Te\'{o}rica, Universidad Aut\'onoma de Madrid, 28049
Cantoblanco, Madrid (Spain)\\ $^{2}$Departamento de Investigaci\'{o}n
B\'{a}sica, CIEMAT, Avda. Complutense 22, 28040, Madrid, (Spain) \\
$^{3}$ INAOE, Luis Enrique Erro 1, Tonanzintla, Puebla 72840, Mexico}

\date{Accepted Received ; in original form }

\pagerange{\pageref{firstpage}--\pageref{lastpage}} \pubyear{2011}

\maketitle\label{firstpage}

\begin{abstract}

We use the self-consistent model technique developed by
Mart\'{\i}n-Manj\'{o}n et al. (2008) that combines the chemical
evolution with stellar population synthesis and photo-ionization
codes, to study the star formation scenarios capable of reproducing
the observed properties of star-forming galaxies.

The comparison of our model results with a database of H{\sc ii}
galaxies shows that the observed spectra and colors of the present
burst and the older underlying population are reproduced by models in
a bursting scenario with star formation efficiency involving close to
20 per cent of the total mass of gas, and inter-burst times longer
than 100\,Myr, and more probably around 1\,Gyr.  Other modes like
gasping and continuous star formation are not favored.

\end{abstract}


\section{Introduction}
H{\sc ii} galaxies can be considered as the strong emission line
subset of the Blue Compact Dwarf galaxies (BCD). Their optical
emission is dominated by strong and narrow emission lines produced by
the interstellar gas ionized by a young and luminous cluster. Their
integrated optical spectra are indistinguishable from a normal
H{\sc ii} region. H{\sc ii} galaxies are gas rich and metal
poor. These facts had led to the early proposal that these systems may
be very young perhaps undergoing their first burst of star formation
(SF) \citep{SS70}. However, many authors have found clear evidences of
the existence of a low surface brightness, old, non-ionizing stellar
population in the majority of these galaxies, as explained
by \citet[][here in after MMDT]{mmm08a} and references therein.  The
existence of an old population plus the relative paucity of H{\sc ii}
galaxies with extreme low metal content are consistent with a scenario
in which H{\sc ii} galaxies are suffering at present a starburst, instead of being old systems with a slow
evolution. Therefore we use the
term {\sl starburst} to denote a violent episode of star formation
where a large number of massive stars have been formed in a small
volume of space and over a time scale of a few million years. Galaxies
called {\sl Starbursts Galaxies} are a very heterogeneous category,
including from BCDs to ULIRGS. We will handle here only H{\sc ii}
galaxies, those dwarf galaxies with active star formation whose effects
actually dominates their UV-optical emission. In these galaxies the
starburst is considered a phase, a process of strong star formation in
terms of intensity and duration.  Nevertheless, the fundamental
details of how the star formation history proceeds in H{\sc ii}
galaxies is still an unresolved problem.

Until now, three basic evolutionary scenarios have been postulated:

\begin{itemize}

\item\textbf{Bursting star formation}: the stars form
in short but intense episodes separated by long quiescent periods of
very low or null activity \citep{davph88,bra98}

\item\textbf{Gasping star formation}: The star
formation takes places as long episodes of SF of moderate intensity
separated by short quiescent periods \citep{tosi91, apa95,recc04}.

\item \textbf{Continuous star formation}: The process
of stellar formation is continuous and of low intensity during the
galaxy life, with superimposed sporadic bursts \citep{leg00}.  

\end{itemize} 

The low metallicities, lack of dust and the optical colors argue in
favour of a bursting star formation with long quiescent periods
\citep[e.g.][]{mar94}. However, inactive star forming periods seem to
rarely occur, and inter-burst stages characterized by low-levels of
star formation are perhaps more plausible than its complete cessation
\citep{lee04}. If this is the case, a continuous low level state of
star formation would dominate during the vast majority of the time
during which many or perhaps most of the stars in BCD galaxies would
be formed.

Chemical evolution models with gasping or bursting star formation
modes seem to be appropriate to describe local dwarf irregulars (dIrr)
\citep{recc04,mar95,tosi91,gall96}, but the chemical evidence alone
cannot differentiate between gasping and bursting scenarios.  The
gasping scenario can be viewed as a transition between the continuous
star formation and the bursting scenario \citep{mar94}. The most
remarkable difference between gasping and bursting scenarios is that
short episodes of star formation enrich the ISM in a time scale of few
tens of Myr while long-lasting episodes enrich gradually the ISM in a
longer time scale, and any further episode of SF does not leave an
appreciable imprint on the chemical evolution. The abundances, which
increase rapidly in bursting models, change more slowly and during a
longer time in gasping models, but with older ages in their stellar
populations \citep{recc01,recc02}.  This picture, however, does not
entirely exclude the possibility that star formation histories appear
to be composed by burst cycles when examined with a high time
resolution, but looks constant when averaged over longer time
scales. It does not exclude either the possibility that the SF
propagates through the galaxy, taking place in independent luminous
short lived H{\sc ii} regions: ``the short timescales associated to
starbursts in dwarf galaxies may be understood as {\sl flickering}
events, small components of a larger starburst in the galaxy''
\citep{mcquinn09}

These observations of stellar populations suggest that it is most likely that the
star formation in dwarf galaxies is sporadic, separated by millions to billions of years, even in isolated systems,
as most of H{\sc ii} are. If the SF of H{\sc ii} galaxies is dominated
by intermittent starburst episodes separated by long inactive time
spans, quiescent blue dwarf galaxies without the dominant starburst
and showing similar properties to those with such intense star
formation events, should be relatively common.  Support to this
framework comes from the fact that blue compact dwarfs (BCD), with
emission lines on average much weaker than these of H{\sc ii}
galaxies, are more common than those ones.  Following this idea
\citet{salmeida08} obtain a relationship between the duration time of
the starburst phase and the quiescent periods, based on the number of
objects of each type of their sample. They find that, if the duration
of the burst phase is 10\,Myr, the time in quiescence must be at least
0.27\,Gyr, implying several mayor starburst episodes along the life of
a BCD galaxy.  Therefore, the diversity in properties exhibited by
dwarfs may be at least partially the consequence of observing them at
different times in their star formation cycles, and the frequency
distributions of galaxies in the various phases could correspond to
the time spent in each of them.

It would be interesting to see which is the most probable star
formation scenario given by the {\sl ab initio} cosmological
hydrodynamical simulations.  However this is difficult since the dwarf
galaxies produced in the existing works are usually poor gas, dwarf
elliptical (dE) or dwarf spheroidals, (dSp) objects
\citep{valcke08,revaz09,sawala10}. Other simulations, such as
\citet{pelu04} used fixed initial conditions for stellar and gas
masses, and this way the mass to halo mass ratio is not a result of
the simulation.  In \citet{stin09} simulations, the Ultraviolet (UV)
background is not included, which probably contributes to a high star
formation efficiency and to large final stellar masses. \cite{mash08}
follow the evolution of a galaxy with a halo mass of 10$^{9}$\,M$_{\odot}$ but only until z=5.  
\cite{gov10} have calculated some
complete hydrodynamical simulations, in which baryonic processes, as 
gas cooling, heating from the cosmic UV field, star formation and 
supernova-driven gas heating, are included with sufficient spatial
resolution, with clumps as low as 10$^{5}$\,M$_{\odot}$ resolved. The
created dwarf galaxies without bulges are similar to the observed
dwarf irregular galaxies. However, besides the authors do not show the
star formation histories, these galaxies show a maximum star formation
rate (SFR) of 0.25\,M$_{\odot} \rm yr^{-1}$ and a present SFR of 
0.01\,M$_{\odot} \rm yr^{-1}$, much lower than observational values for H{\sc
ii} galaxies, and not comparable to these objects that we will try to
model here.  Only \cite{nag10} has recently presented some star
formation histories, obtained from cosmological hydrodynamical
simulations for dwarf galaxies, with enough spatial resolution for
gravitational masses of ~10$^{9}$\,M$_{\odot}$. Their star formation
histories seem to be sporadic, and the stars continue forming sporadically even
at late times. However, the simulations are still having some problems to
reproduce the adequate number of galaxies and the author claims that
more work is still necessary. As \cite{sawala11} explains, all
dwarf galaxies (~10$^{10}$\,M$_{\odot}$) formed in the current
hydrodynamical simulations are more than an order of magnitude more
luminous than expected for these masses. In any case they are much more massive than 
BCDs and H{\sc ii} galaxies that we want modeling.

For what refers to other models existing in the literature, there are
not models for the study of BCD or HII galaxies in the way we want to
use them. A number of models have computed purely chemical evolution
models for BCDs or dwarf irregular galaxies \citep[][ and many
others]{chi82,mar94,recc02,recc03,shi06,recc07}. Most of them assume
that the star formation occurs in bursts and include the effects of
galactic winds and/or gas infall. However, they limit the study to the
evolution of N and O abundances and/or to the luminosity-metallicity
relation. \cite{mouh02} and \cite{vaz03} used the information coming
from the chemical evolution models from \cite{car02} to perform the
next step and combine chemical and spectral evolution for irregular
galaxies. These models, however, exclude the early stages of
evolution, i.e. during the nebular phase when most massive stars
dominate the energy output budget. \cite{kru91,lin99,vaz97} or
\cite{hoek00} compute chemical and photometric evolution models in a
way more or less consistent for starburst, spiral, early types or low
surface brightness galaxies, respectively, any of them applied to H{\sc
ii} galaxies. Other works, in turn, are focused on the ionized gas
properties and make models using Single Stellar Populations (SSPs) spectral
energy distributions (SEDs) of a given metallicity for applying a
photo-ionization code to obtain the emission lines and studying
diagnostic diagrams and/or abundances obtained by empirical
calibrations \citep{sta96,sta01,sta03,moy01,dop06,mmanjon10}. Most of them
ignore the star formation history of the galaxy and not take into
account the underlying stellar population, so these studies are only
valid for the study of the current stellar generations of the galaxy.
Summarizing a code combining chemical, evolutionary synthesis and 
photo-ionization models has not been applied for the analysis of BCDs and
H{\sc ii} galaxies, with the exception of the ours (MMDT),
but only to elliptical massive galaxies \citep{bre94}.

The aim of this work is to critically analyze the possibility of dwarf galaxies undergo recurrent phases of star formation by comparing the predictions of our self-consistent evolutionary models with the
most evolution sensitive observed parameters, i.e. stellar continuum
colors, emission line equivalent widths and chemical composition in a
large sample of BCD galaxies.

We have used an updated version of our models from MMDT, developed to
predict the main characteristics of H{\sc ii} galaxies, such as the
emission lines, continuum color, continuum plus the contribution of
emission lines colors, equivalent widths and chemical abundances.  Our
code uses the chemical evolution model results for the computation of
the SEDs, which are, in turn, used as the ionizing sources for a
photo-ionization code.  In MMDT we studied the viability of this tool,
showing its adequacy in the reproduction of the data defined by the
stellar populations (colors, equivalent width of absorption lines or
spectral indices, spectral energy distributions..), which define the
time evolution, as well as the characteristics of the gas phase
(emission lines, equivalent widths, elemental abundances, gas
densities), which defines the present time state of the galaxy. This
is done in a self-consistent way, that is, using the same assumptions
regarding stellar evolution, model stellar atmospheres and
nucleosynthesis, and using a realistic age-metallicity relation.

 One of the most important results obtained from our previous work was that observational data are reproduced
only if the mass involved in the last burst is much smaller than the
mass of the old underlying stellar population.  Now, once checked the
possibilities of our tool, we try to wide the number of models,
changing the star formation scenarios to see if data are reproduced
with more or less success for some of them. 

It is a widely assumed idea that starburst galaxies are strongly
affected by gas infall and outflow and that chemical abundances and
star formation cannot modeled without these issues. However, the
existence of outflows in dwarf starburst galaxies is by no means a
settled issue; while winds able to escape the galaxy have been found
in two prototypical starburst galaxies like NGC1569 and NGC1705, recent
works about mass loss, for example \cite{eymeren07,eymeren09,eymeren09b, eymeren10}, find
that ionized and neutral hydrogen expansion velocities measured
are, in all cases, too low to allow the gas to escape from the
gravitational potential of their studied galaxies some of which, like
NGC2363 or NGC4861, can also be considered as {\sl prototypical}. In
fact, according to \cite{bomans07}:''While the observational support for
the presence of galactic winds in massive galaxies and gas-rich
mergers is quite strong, the case for galactic winds in dwarf galaxies
is much weaker''. 

The need of winds is also related with the low oxygen abundance found
for dwarf galaxies for their gas fractions, and, in this case, both
infall and outflow can help. However, galaxies loosing a large
fraction of their gas should show red colors, contrary to what is
observed. In fact, a good fraction of the galaxies in the sample of
\cite{vzee06} can be reproduced with closed box models. Furthermore,
it is necessary to take into account the N/O vs O/H relationship for
dwarf galaxies. Selective winds models as \cite{mar94,bra98} have
recognized problems in reproducing the observed N/O ratios, and the
work by \cite{lar01} demonstrates that the observed N/O ratios, as
well as their dispersion, can be reproduced with closed box models
while selective winds can be ruled out. Actually, \cite{mol06} showed that it is mainly the combination of the different time
scales in nitrogen production by stars of different masses and the
galaxy star formation histories as proceeding from different star
formation efficiencies, what establishes the N/O ratio.

Therefore, since there is enough proof to conclude that
infall and/or outflows are not necessary to reproduce the general
trends of H{\sc ii} galaxies, galactic supernova-driven winds are not
included in our models. This scenario is also in agreement with models
from \cite{tassis08}, who find that scaling relations of dwarf
galaxies may be reproduced by simulated galaxies without
supernova-driven outflows.

For the present work, new and updated theoretical codes have been
applied for the computation of the models. The main changes correspond
to the evolutionary synthesis models, now from \citet[][hereinafter
MGVB09]{mgvb09} instead \cite{gbd95}. As explained before, to guarantee 
the self-consistency of the approach we have used the same assumptions in the stellar
evolution, model stellar atmospheres and nucleosynthesis parts of the
code.  The use of these new models allows to be more consistent than
in our previous work MMDT, since there the IMF used in the chemical
evolution models \citep{fer90} and in evolutionary synthesis models
\citep{sal55} was not the same. The new models MGVB allow to us use
the same IMF for chemical and evolutionary synthesis (photometrical)
calculations. Moreover, with the previous models from \cite{gbd95} we
had also a problem for the lowest metallicities stellar populations
since models for the youngest ages of low metallicities were not
available. Now we have models for the same number of ages for all
metallicities. To include the lowest metallicities of the youngest ages 
change appreciably the colors of the stellar populations and allows to
calculate emission lines for very low metallicity regions, as it is
shown in \citet{mmanjon10}. The contribution of the emission lines is
quite different than expected from the simple extrapolation from other
metallicities, also modifying the final results of colors.
 
The data shown in the graphs have been extracted mainly from two main sources. 
First, the compilation from \citet{hoy06} which provides emission
line measurements, corrected for extinction, published for local 
H{\sc ii} galaxies.  The sample comprises 450 objects and constitutes a
large sample of local H{\sc ii} galaxies with good-quality spectroscopic
data. The sample is rather inhomogeneous in nature, since the data
proceed from different instrumental setups, observing conditions and
reduction procedures, but have been analysed in a uniform way. Data
for these sample objects include the emission line intensities of:
[OII]$\lambda\lambda$ 3727,29 \AA\, [OIII]$\lambda\lambda$ 4959,5007
\AA\, and [NII]$\lambda\lambda$ 6548,84 \AA\, all of them relative to
H$\beta$, and the equivalent widths of the [OII] and [OIII] emission
lines -- EW([OII]) EW([OIII]) --, and the H$\beta$ line, EW(H$\beta$).
As a second source, we have used the metal poor galaxy data from the
Data Release 3 of Sloan Digital Sky Survey, taken from
\citet{izo06}. The Sloan Digital Sky Survey \citep[][]{yor00}
constitutes a large data base of galaxies with well defined selection
criteria and observed in a homogeneous way.  The SDSS DR3 \citep{aba05}
provides spectra in the wavelength range from 3800 to 9300 \AA\ for
$\sim$ 530000 galaxies, quasars and stars.  \citet{izo06} extracted
$\sim$ 2700 spectra of non-active galaxies with the
[OIII]$\lambda$4363 \AA\ emission detected above 1$\sigma$ level. This
initial sample was further restricted to the objects with an observed
flux in the H$\beta$ emission line larger than 10$^{-14}$ erg s$^{-1}$ cm$^{-2}$ 
and for which accurate abundances could be derived. They have
also excluded all galaxies with both [OIII]$\lambda$4959/H$\beta$ $<$ 0.7 
and [OII]$\lambda$3727/H$\beta$ $>$ 1.0.  Applying all these
selection criteria, they obtain a sample of $\sim$ 310 SDSS
objects. Data for these sample objects include the emission line
intensities of: [OIII]$\lambda$4959,5007 \AA\ and [NII]$\lambda$6584
relative to H$\beta$ and the equivalent width of H$\beta$.  They also
include the intensity of the [OII] $\lambda\lambda$ 3727,29 \AA\
emission line for the lowest redshift objects.

In the next section an explanation of the theoretical star-bursting models
is made. The section 3 presents the results, the meaning of each input
parameter and the implication of their variation over the results of
the models. We will discuss the influence of these parameters in the
reproduction of the observable characteristics of star-forming galaxies,
their impact over possible star-formation scenarios and the connection among
different type of dwarf galaxies in section 4. Finally, a summary
and the conclusions of this work are presented in section 5. All
tables of the models will be available in electronic format.

\section{Theoretical Models}

The viability of our model technique to reproduce the observable
characteristics of H{\sc ii} galaxies was discussed in MMDT: In a
first stage the chemical evolution model is computed by using certain
input parameters to obtain the star formation history and the gas and
stars chemical abundances. Secondly, the evolutionary population
synthesis code is applied to obtain the SEDs corresponding to each
time step of the chemical evolution.  Finally, the ionizing part of
the SED and the resulting abundances are used as an input to the
photo-ionization code to compute the time evolution of the emission
lines of the ionized gas.

\subsection{Chemical evolution}

The chemical evolution is computed with a simplified version of the
classical {\sl multiphase chemical evolution model} from
\citet{fer94,mol96,mol05}. In our version of the code there is only one
region (without the two zones halo and disk) \footnote{It implies that
there is no infall of gas over a disc, as in the spiral and irregular
galaxies models} with a given mass of gas which form stars (in this
case we do not consider the phase of formation of molecular clouds as
in the classical multiphase models).  We have run models considering
the star formation as a set of successive bursts followed by quiescent
periods in a region with a total mass of gas of $\rm 10^{8}\,M_{\odot}$. 
In each burst a given amount of gas is consumed to form
stars, and this process comes defined by the star formation efficiency. The code
solves the chemical evolution
equations to obtain, in each time step, the abundances of 15 elements:
H, D, $\rm ^{3}He$, $\rm ^{4}He$, C, $\rm ^{13}C$, O, N, Ne, Mg, Si,
S, Ca, Fe, and \textit{nr} (where \textit{nr} are the isotopes of the
neutron rich elements, synthesized from $\rm ^{12}C$, $\rm ^{13}C$,
$^{14}$N and $\rm ^{16}O$ inside the CO core). The stellar yields are
those from \citet{ww95} for massive stars, $\rm M > 8\,M_{\odot}$, and
those from \citet{gav05} for low and intermediate mass stars. The
supernova Ia yields used proceed from \citet{iwa99}. The initial mass
function (IMF) is taken from \citet{fer90} with a range between 0.15
and 100 $\rm\,M_{\odot}$.  More recent IMFs such as \citet{kro02} or
\citet{cha03} there exist but, for chemical evolution models, it is
necessary to use a combination IMF+stellar yields sets calibrated with
Milky Way Galaxy data, that is, able to reproduce the elemental
abundances and gas and star densities as observed. This calibration is
not still done for those IMFs, although an update of models taking
these IMFs and some new stellar yield sets will be shown in a next
future (Molla et al. in preparation).  We have taken time steps of
$\delta$t$\sim$ 0.7\,Myr\footnote{The time step is chosen to include
the fastest evolutionary phases of the most massive stars} from the
initial time, $t=0$, up to the final one, $t= 13.2$\,Gyr.  At each time
step, the star formation rate and the mass in each phase -- low mass,
massive stars and remnants, total mass in stars created, and mass of
gas-- are also obtained.

\subsection{Evolutionary synthesis}

The SEDS  are taken from the PopStar
evolutionary synthesis models by MGVB09. Isochrones are an update from
those from \citet{bgs98} for 6 different metallicities: $\rm Z=
0.0001$, 0.0004, 0.004, 0.008, 0.02 and 0.05.  The very low
metallicity model of $\rm Z=0.0001$ had not been included before in
similar works. The age coverage is from $\log{t}=$5.00 to 10.30 with a
variable time resolution of $\Delta(\log{t})= 0.01$ in the youngest
stellar ages.
We have used the Ferrini IMF results \citep{fer90} with mass limits between 
0.15 and 100\,M$_{\odot}$ in order to avoid any
inconsistency with the chemical evolution code, which also uses this IMF.

To calculate the SED integrated over the whole history of the galaxy,
SEDs of SSPs  with the corresponding metallicity and age must be convolved with
the star formation history (SFH):
\begin{equation}
L_{\lambda}(t)=\int^{t}_{0}S_{\lambda}(\tau,Z(t'))\Psi(t')dt'
\label{sp}
\end{equation}
where $\tau=t-t'$ is the age of the stellar population created in a
time $t'$ and $S_{\lambda}$ is the SED for each SSP of age $\tau$
and metallicity Z reached in that time $t'$.  A SED from the SSP
library, $S_{\lambda}$, must be assigned to each time step according
to its corresponding age and metallicity. Taking into account the SFH,
$\Psi(t)$, and the age-metallicity relation, $Z(t)$, obtained from the
chemical evolution model, we know the age and metallicity assigned to
each time step or stellar generation. For each one of them we have
chosen the SED closest in age among those available in the grid of
PopStar. However, in our models the metallicity changes continuously
while the available SEDs of the library are computed only for 6
possible values.  Therefore, we have interpolated logarithmically
between the two SSP of the same age $\tau$ and closest in
metallicities to $Z(t)$ to obtain the corresponding
$S_{\lambda}(\tau,Z(t'))$.  The final result is the total luminosity
at each wavelength $\lambda$,  L$_{\lambda} \,(\rm erg.s{-1}.\AA^{-1}$),
corresponding to the whole stellar population, including the ionizing
continuum proceeding from the last formed stellar population.

\subsection{Photo-ionization calculations}

To compute the photo-ionization models, we have used {\sc cloudy}
\citep [version c06.02][]{fer98}\footnote{In order to maintain the
consistency with MMDT, we have used the same
Cloudy version for the present work. The differences between the used
version and the laters are new features and new possibilities for the
modelization, however, there are not significative differences in our
model results.}.  The gas is assumed to be spherically symmetric
around a point source of radiation and the pressure or density in the
gas is imposed by external conditions. This allows a plane-parallel
geometry treatment in which the gas may be regarded as a thin shell. A
closed geometry has been taken for the calculations. All the photons
which escape from the illuminated face of the cloud towards the star,
go on to strike the other side of the nebula, ensuring the case B of
recombination and the approximation on the spot.  The number of
ionizing photons, Q(H), striking the illuminated face of the cloud,
have been calculated directly from the total resulting SED of the
models.

We derive the radius of the modelled region from the mechanical
energy from massive stars with strong winds (taken from MGVB09), instead of using the radius required to 
maintain a constant density of stellar mass through the successive
bursts, as we did in MMDT.  \citet{cmw75} demonstrated that
an early-type star with a strong stellar wind can blow out a large
cavity or {\sl bubble} in the surrounding gas, if it is assumed to be
compressed into a thin spherical shell. The wind-driven shell begins
to evolve with an initial phase of free expansion followed by an
adiabatic expansion phase, and then the material collapses into a
thin, cold shell as a result of radiative cooling. At this stage the
gas traps the ionization front and the radiative phase begins. In this
phase the ionizing photons are absorbed and the region cools via
emission in the Balmer lines. In this process, the radius of the outer
shock, R$_{s}$, evolves as:
\begin{equation}
R_{s}=1.6(L_{mec}/n)^{1/5} t^{3/5} \,\rm{pc}
\label{Rs}
\end{equation}
where $L_{mec}$ is the total injected mechanical energy (SN and
stellar winds) per unit time in units of 10$^{36}$\,ergs s$^{-1}$, $n$
is the interstellar medium density in units of cm$^{-3}$, and $t$ is
the age of the shell in units of 10$^{4}$\,yr. Since we use the value of the energy
at each time-step, this radius represents the instantaneous size of
the region obtained by adding the winds and SNe from the previous age
to this age. Therefore  L$_{mec}$  is this energy divided by the time step.
Then, the ionized gas is assumed to be located in a thin spherical
shell at that distance R$_{s}$ from the ionizing source. This approach
has the advantage of eliminating the ionization parameter as a free
variable in the models since now it is computed from the physical
parameters of the evolving young cluster.

The ionized gas  abundances are assumed to be those reached at the end of the
starburst previous to the current one. Fifteen element abundances have
been introduced in the photo-ionization code: He, C, N, O, Ne, Na, Mg,
Al, Si, S, Ar, Ca, Fe and Ni, obtained from the chemical evolution
model, except for Na, Ar and Ni which are not computed in the model
and are scaled to the solar ratio \citep{asp05}. The models assume
that the nebula is ionization bounded and no dust has been
included in the chemical evolution models neither in the photo-ionization calculations. 
However dust grains mixed with the ionized gas have been partially
taken into account, since we have included the depletion in refractory
elements (Si, Fe, CA, Si, Mg) taken from \citet{gar95}. The
grains can affect the absorption of the \textit{UV} photons and decrease
the electronic temperature. The density has been assumed constant for simplicity
and equal to 100\,cm$^{-3}$,  which is appropriate for
modeling H{\sc ii} galaxies \citep{hag08} and large circumnuclear
H{\sc ii} regions \citep{gv97, diaz07}, frequently found around the
nuclei of starbursts and AGNs. Although the constant density
hypothesis is probably not realistic, it can be considered
representative when the integrated spectrum of the nebula is analyzed.

The shape of the ionizing continuum is
defined by the pair of values ($\nu$(Ryd), log$\nu L_{\nu}$)
(Eq.~\ref{sp}) obtained from the ionizing spectrum given directly by
the evolutionary synthesis code.

\subsection{Input parameters}

Taking into account the simplified version used for our chemical
evolution models, the free input parameters related with the infall
time-scale and the molecular cloud formation disappear. The total mass
of the initial gas is within the region from the initial time. The star
formation rate is zero except during the bursts. We need to define the
intensity of these bursts, and the time elapsed between them.

Each model is therefore characterized by three input parameters:

\begin{itemize}
\item \textbf{1 - The Initial efficiency (\textit{$\epsilon$})}: It is the
amount of gas consumed to form stars in the first burst of star
formation, that is, $\Psi(t)=\frac{dM_{s}}{dt}=  \epsilon M_{g}$

We present here models computed with 2 values of $\epsilon$:

\begin{enumerate}

\item High efficiency models: The first burst of star formation
involves 33  per cent of the total initial mass of gas
(33$\times$10$^{6}$\,M$_{\odot}$),

\item Low efficiency models: The first burst of star formation
involves 10 per cent of the total initial mass of gas
(10$\times$10$^{6}$\,M$_{\odot}$).
\end{enumerate} 

\item \textbf{2 - Attenuation}: The star formation efficiency of the
following bursts is attenuated in two different ways:
\begin{enumerate}

\item By a factor which changes with the number of the burst,
\textit{n}, according to the expression:

\[
\Psi_{n}=(\frac{1}{n})\cdot\Psi_{0} 
\]

This attenuation type corresponds to a soft attenuation model (hereinafter SAM).

\item By a constant factor, \textit{k$^{(n-1)}$}, according
to the expression:

\[
\Psi_{n}=\Psi_{0}\cdot k^{(n-1)}
\]

corresponding to a high attenuation model (hereinafter HAM). In this last
expression \textit{k} is the attenuation factor which, in this case
\footnote{It may be higher than 1 in models with increasing
efficiency; we have computed some of these models, however, their
results are in clear disagreement with most observations.}, takes values
from 0 to 1. The lower the \textit{k} value, the stronger the
attenuation, the burst being less efficient each time; the higher the
\textit{k} value, the weaker the attenuation, and then the stronger
the bursts. We only show models with $k=0.65$

\end{enumerate} 

\item \textbf{3 - Time between bursts (\textbf{\textit{$\Delta$t}})}:
Every burst takes place instantaneously and it is followed by quiet
periods, whose duration can change. For this work we have taken
$\Delta$t= 1.3\,Gyr for the inter-burst time, that is, one burst every
1.3\,Gyr as the generic case, although we will also show some models with
$\Delta t=0.1$\,Gyr and $\Delta t=0.05$\,Gyr in order to compare the
different results.
\end{itemize} 

The models we show here are a selection of the ones calculated and
described more widely in Mart\'{\i}n-Manj\'{o}n, Moll\'{a} \&
L\'{o}pez-S\'{a}nchez, 2012 (in preparation). In that work, the results of
20 models are compared with data for generic and particular low mass
dwarf galaxies.  We have selected the most relevant of them in order
to show the effects of the input parameter variation on
the resulting evolutionary history of each galaxy. The input
parameters of these six models are shown in Table~\ref{inputs}. 

\begin{table}
\caption{Input parameters for the theoretical models. In column 1 it is defined the type of model according to the attenuation:
soft (SAM) or high attenuation (HAM); The second column gives the
parameter $k$ which defines the attenuation; column 3 is the star
formation initial efficiency $\epsilon$, and column 4 shows the time
between burst $\Delta t$ in\,Gyr.}
\begin{tabular}{ccccc}
\hline
Num. & Attenuation  & $k$ & $\epsilon$ &  $\Delta t$\\
of model & type   &      &      &Gyr \\
\hline
1 & SAM    &   --   &  0.10 &  1.30  \\ 
2 & SAM    &   --   &  0.33 &  1.30  \\ 
3 & HAM    &  0.65  &  0.10 &  1.30  \\  
4 & HAM    &  0.65  &  0.33 &  1.30  \\ 
5 & HAM    &  0.65  &  0.33 &  0.10  \\   
6 & HAM    &  0.65  &  0.33 &  0.05  \\ 
\hline
\label{inputs}
\end{tabular}
\end{table}

\section{Results}

\begin{figure}
\resizebox{\hsize}{!}{\includegraphics[angle=0]{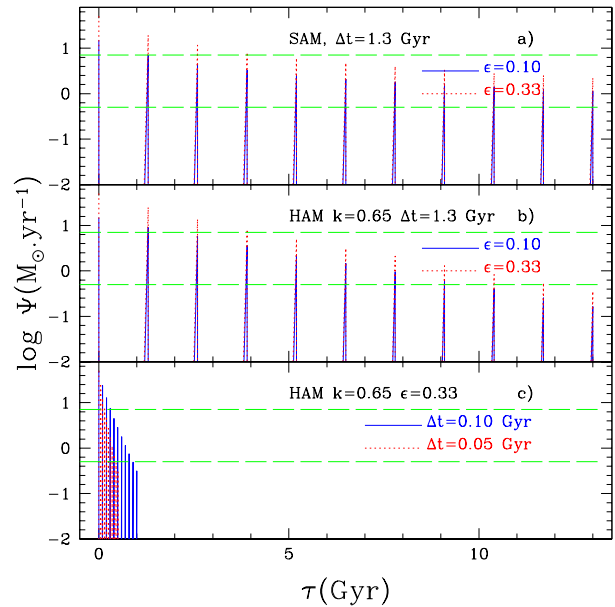}}
\caption{SFR of the models of the table~\ref{inputs} as labelled.  In
a) we show the SAM (1 and 2), in b) two HAMs (3 and 4), with the 2
same efficiencies as in a).  For these 4 models the time between
bursts is $\Delta t=1.3$\,Gyr.  In c) we show two models (5 and 6)
with shorter time between bursts, $\Delta t=0.1$ and 0.05\,Gyr.  The
dashed (green) lines define the upper and lower limits to the SFR
estimated for BCD and/or H{\sc ii} galaxies by \citet{hoy04}.}
\label{fig1}
\end{figure}

We will show here the main results obtained with the selected models.
The \textbf{efficiency} determines the initial star formation rate and
the initial metallicity of the gas. We have plotted in Fig.~\ref{fig1}
the SFR for the 6 models.
The first burst is strong in all panels,
while the subsequent ones are less intense due to the decrease of the
available gas to form stars and the attenuation. In a) and b) the last
time computed is $~ 13$\,Gyr while in c) and d) the final time is around 1.3
and 0.6\,Gyr respectively. In all panels we show the observational limits
as dashed (green) lines. Taking these limits into account, the dwarf
galaxies may suffer 11 bursts for SAM, or 8-9 for HAM
since later bursts show rates lower than observed.

\begin{figure}
\resizebox{\hsize}{!}{\includegraphics[angle=0]{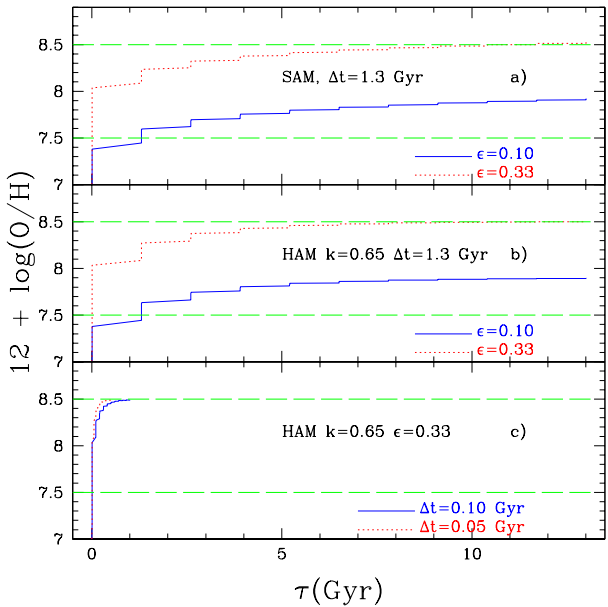}} 
\caption{Evolution of oxygen abundance for same models
and with the same line code as in Fig.~\ref{fig1}.
The dashed (green) lines define the observational data range}
\label{fig2}
\end{figure}

The oxygen abundances for these same models are shown in
Fig.~\ref{fig2}. In this plot we can see that the two efficiencies
chosen, 33 per cent and 10 per cent, give the upper and lower limits
respectively for the observed oxygen abundance range in this type of
galaxies. The models with the same initial efficiency and different
attenuation type, are very similar, and if $\Delta t$ is shorter, the 
low oxygen abundance limit is reached very quickly.

\begin{figure}
\resizebox{\hsize}{!}{\includegraphics[angle=-90]{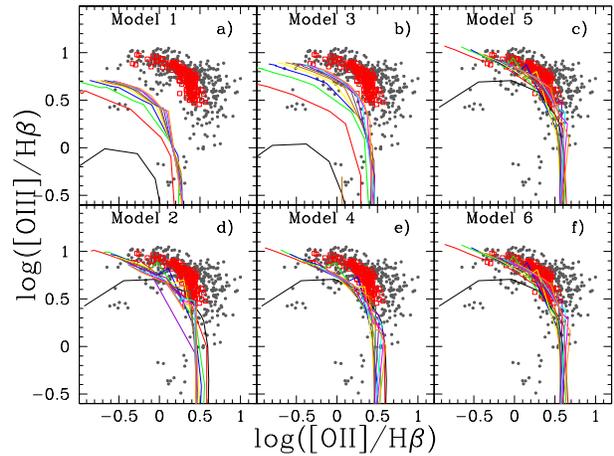}}
\caption{Diagnostic diagram for the computed models involving
[OIII]$\lambda\lambda$5007,4959/H${\beta}$ {\sl vs.}
[OII]$\lambda$3727/H${\beta}$ of low 
(top panels) and high (bottom panels) initial efficiency models
corresponding to the SAM at the left and to the HAM at the right and
center panels. The different colored lines represent each burst, from
the first one occurred at t=0\,Gyr (black line) to the last one at
t=13\,Gyr in models 1 to 4, at t=1.3\,Gyr in model 5 and t=0.6\,Gyr in
model 6 (orange line). Observational data are from \citet{hoy06}
--open red squares-- and \citet{izo06} --grey dots--.  The error of
the observational data are less than 1$\%$ of the line intensity, and
therefore , they are not included in the figure.}
\label{fig3}
\end{figure}

\begin{figure}
\resizebox{\hsize}{!}{\includegraphics[angle=-90]{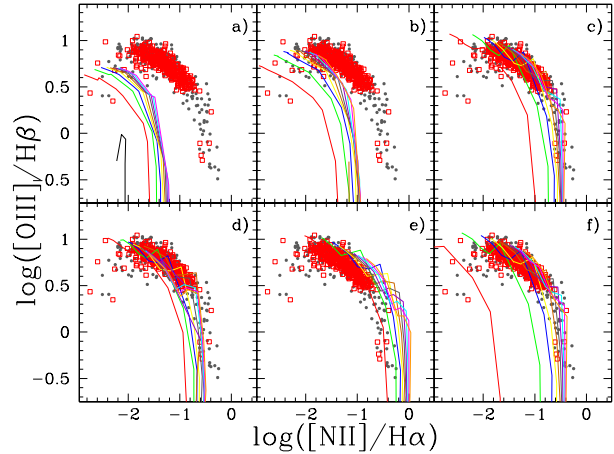}}
\caption{Diagnostic diagram for the computed models involving
[OIII]$\lambda\lambda$5007,4959/H${\beta}$ {\sl vs.}
[NII]$\lambda$6584/H${\alpha}$. The different
colored lines and dots has the same meaning as Fig~\ref{fig3}.
The error of the observational data are less than 1$\%$ of the line intensity, and
therefore , they are not included  in the figure.}
\label{fig4}
\end{figure}

Figs.~\ref{fig3} and \ref{fig4} show diagnostic diagrams involving
the ratios of intense emission lines.
The number of the model (see Table 1) is given in each panel.  
The higher the efficiency, the more ionizing photons produced and the
higher ionization parameter, leading to a higher excitation of the gas. The
low efficiency models cover
the region of the diagram occupied by the young and less metallic
galaxies (high [OIII]$\lambda\lambda$5007,4959/H${\beta}$ , low
[OII]$\lambda$3727/H${\beta}$ and low [NII]$\lambda$6584/H${\alpha}$),
as expected. The high initial efficiency models reproduce high
excitation and high abundance galaxies, with high
[OIII]$\lambda\lambda$5007,4959/H${\beta}$ and high
[NII]$\lambda$6584/H${\alpha}$.  The ionization degree is
driven by the efficiency of the bursts, not by the attenuation mode or
attenuation factor.Thus, the models with different attenuation
modes and the same initial efficiency do not show noteworthy
differences, since it is the current burst of star formation which
produces the observed emission lines, and the underlying population
does not affect the ionization parameter.

The attenuation sets the SFR of the successive bursts and determines
the contribution of the underlying population.  A higher attenuation
implies a larger contribution of previous bursts to the total
SED. Therefore, it is important to see the attenuation factor effect
on the the evolution of the broad-band continuum colors.

\begin{figure*}
\resizebox{\hsize}{!}{\includegraphics{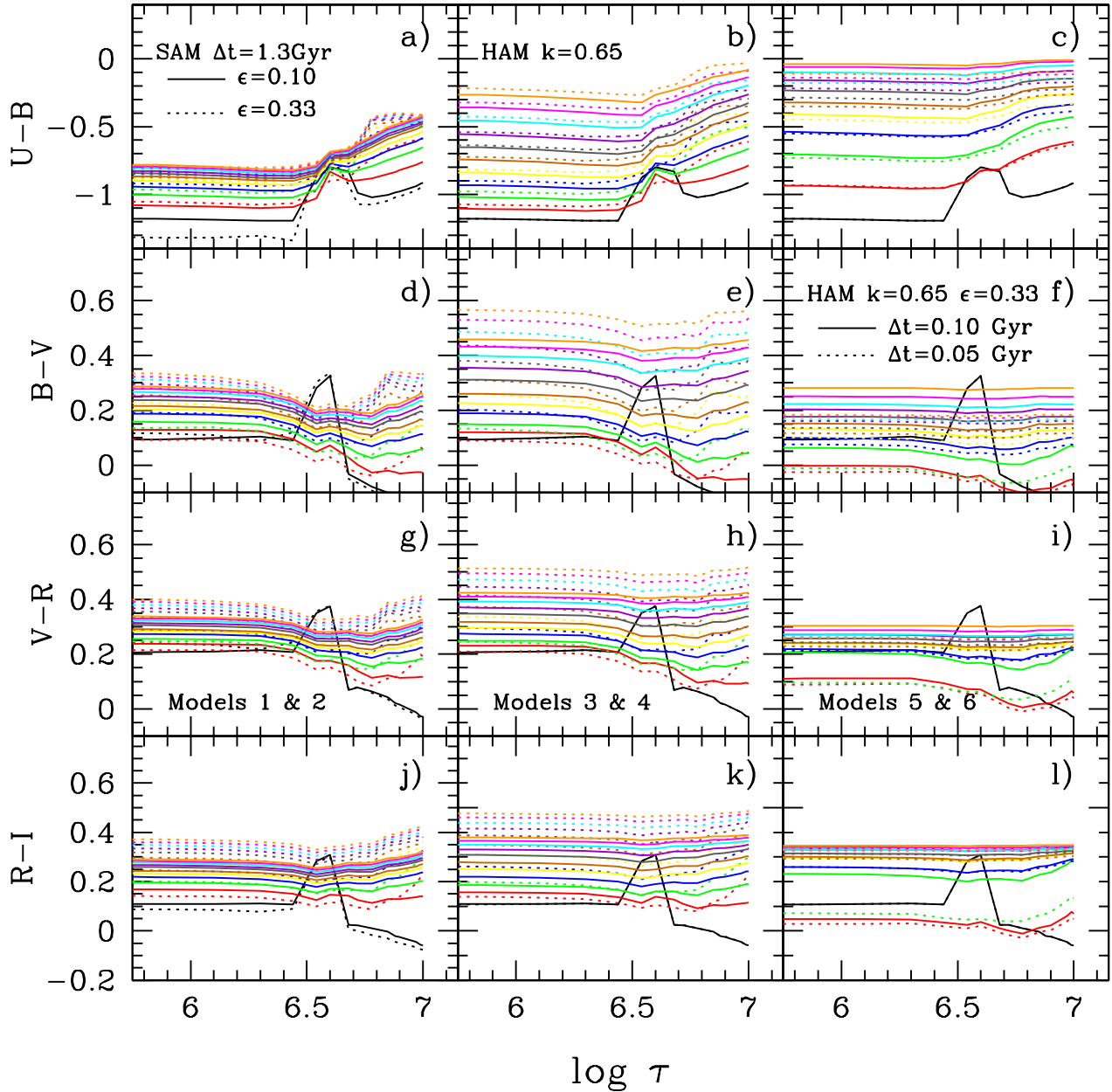}}
\caption{ Evolution of the continuum colors, U-B, B-V, V-R and R-I for
each burst along 10\,Myr. The SAM are plotted on the left, the HAM
with $\Delta t=1.3\,Gyr$ on the intermediate panel and HAM with
$\Delta t=0.1$, and $\Delta t=0.05$\,Gyr, respectively, on the right
of the diagram. All panels include the low and the high efficiency
models as dashed and solid lines, respectively, excepting for the
models on the right panels, wich are showing both high efficiency
models with different inter-burst times, 0.1 and 0.05\,Gyr as solid
and dashed lines, respectively. Each burst is represented with a
different colour: from the first burst (black line) to the last one
(orange line).}
\label{fig5}
\end{figure*}

\begin{figure*}
\resizebox{\hsize}{!}{\includegraphics{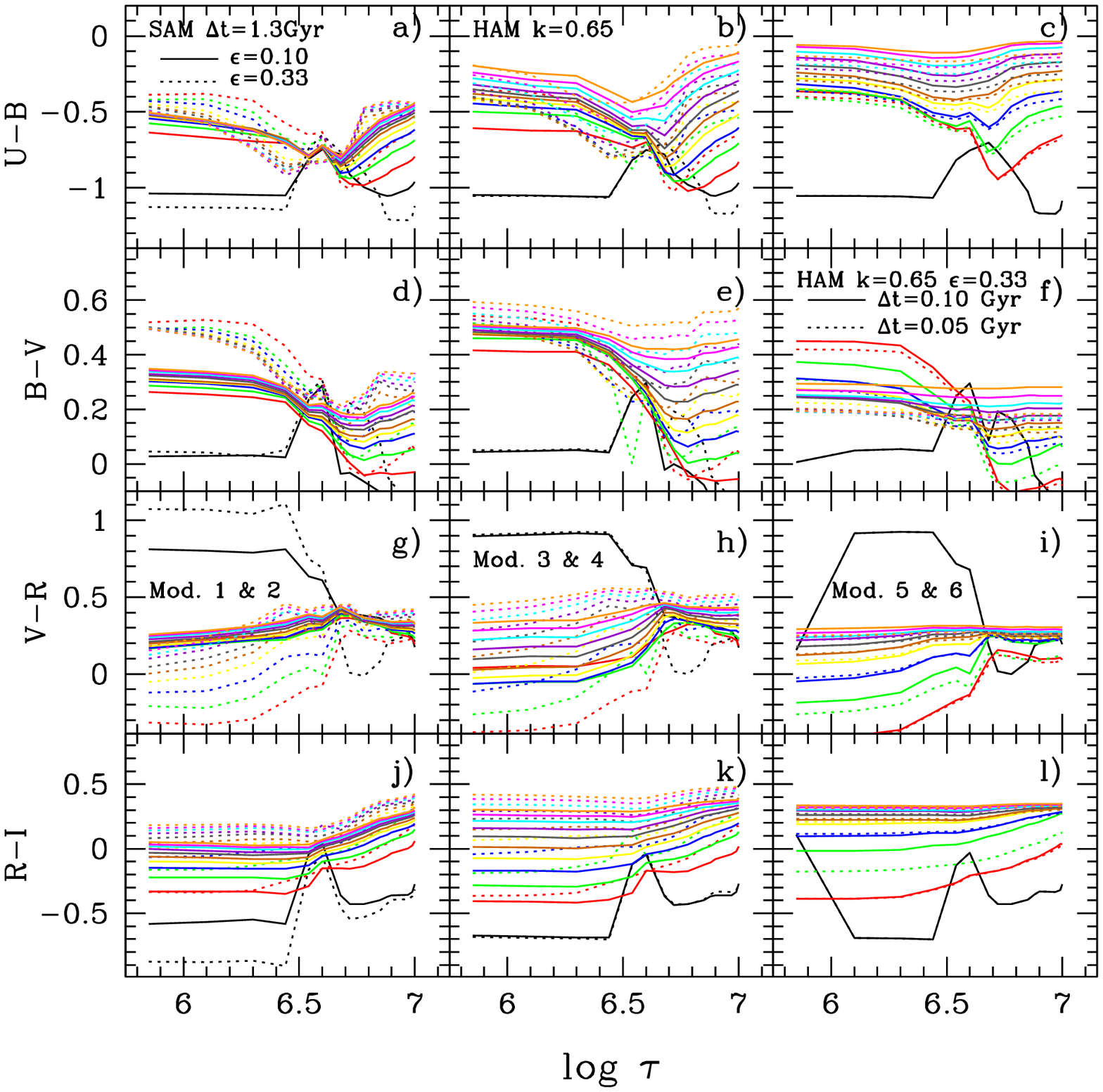}}
\caption{ Evolution of the continuum colors, U-B, B-V, V-R and R-I for each
 burst along 10\,Myr by including the contribution of the emission lines. 
 Each model and each burst are represented with a different colour and line type similar to Figure \ref{fig5}.}  
\label{fig6}
\end{figure*}

In Fig.~\ref{fig5} we show the evolution of the modeled continuum
colors for each burst of star formation. 
The inclusion of nebular emission continuum in the
computation of the SSPs reddens the colors of very young populations
significantly, mainly at low metallicity as explained in MGVB09.

The HAM cases, compared with SAM, have a higher contribution
from the non ionizing underlying continuum, which makes the colors
redder for subsequent bursts and the young and blue population
features disappear. Meanwhile, SAM maintains blue colors,
characteristic of the current burst of star formation.

In Fig.~\ref{fig6} the evolution of the colors is shown including the
emission lines contribution to the wide band filters. 
We have done so taking into account the strongest emission
lines that contribute to the color in each broad-band spectral
interval at redshift zero. These are mainly [OII]$\lambda\lambda$3727 in
\textit{U}, H${\beta}$ in \textit{B}, [OIII]$\lambda\lambda$5007,4959
in \textit{V}, H$\alpha$ in \textit{R} and
[SIII]$\lambda\lambda$9069,9532 in \textit{I}.  In that case
the more intense the last burst, the larger the change in the colors,
so the dotted lines are more similar to colors calculated only with
the stellar populations than those shown by solid lines for higher
efficiencies. On other hand, not all bands are equally modified at the same time.

A fine tuning of the attenuation factor can be obtained
from the evolution of the broad-band continuum colors versus the
equivalent width of H${\beta}$, EW(H${\beta}$). These are important
observations which give also an effective method to uncover the
presence of old underlying stellar populations \citep{ter04}.

\begin{figure*}
\begin{center}
\resizebox{\hsize}{!}{\includegraphics[clip,angle=0]{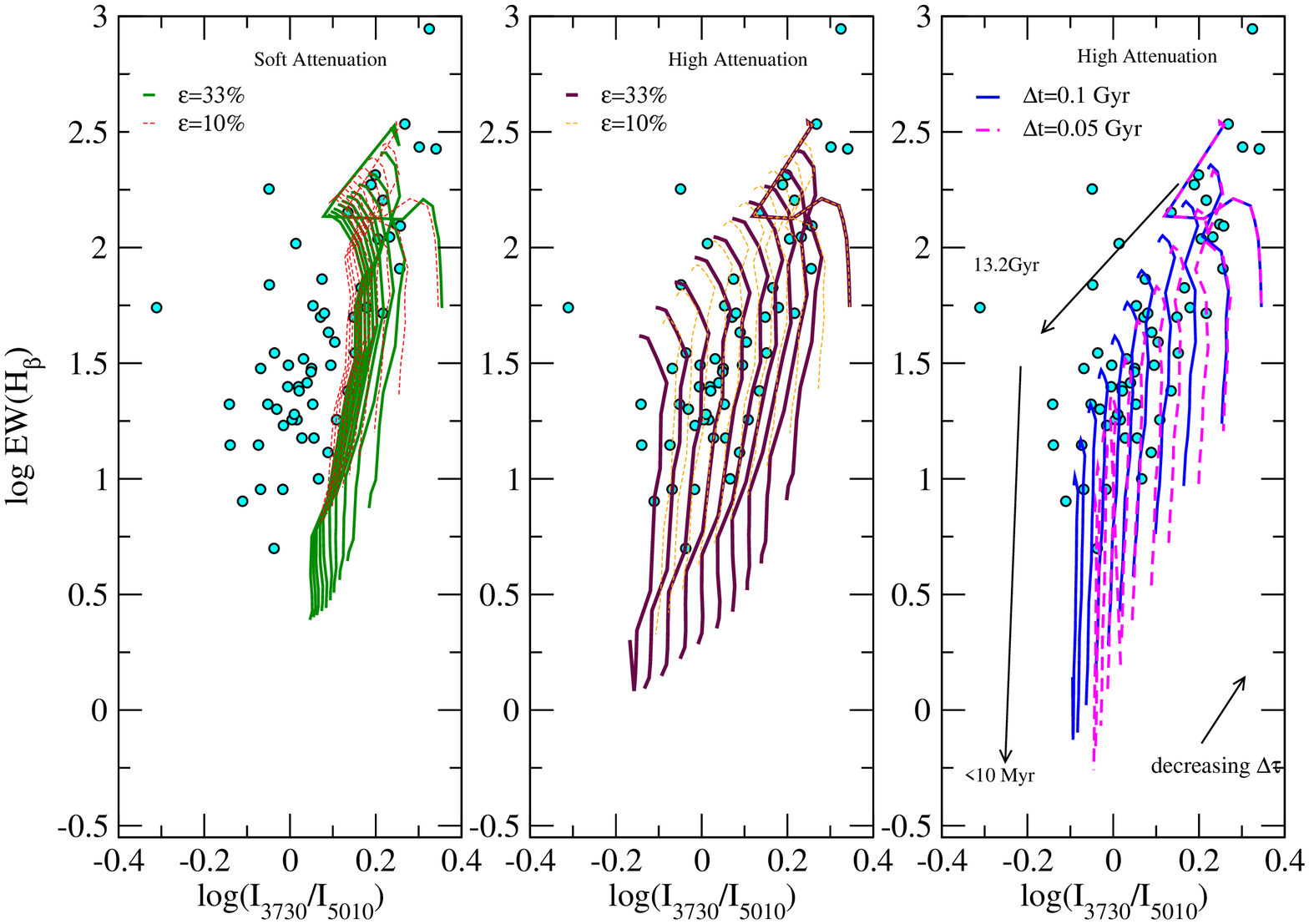}}
\caption{ EW(H${\beta}$) vs. log(I$_{3730}$/I$_{5010}$) compared with
observational data from \citet{hoy06}, \citet{ter91} and
\citet{sal95}, for different attenuation parameters: SAM (left panel),
HAM (central panel), and including both, high and low, initial
efficiencies, shown as solid and dashed lines respectively . The third
panel, on the right, corresponds to high efficiency HAM with different
inter-burst time, $\Delta t=0.1$\,Gyr (solid blue lines) and $\Delta
t=0.05$\,Gyr (dashed magenta lines).}
\label{fig7}
\end{center}
\end{figure*}

The comparison of models and observations shows that in most
star-forming dwarf galaxies, SSPs are unable to reproduce the reddest
colors shown by the data with the lowest EW(H${\beta}$) values. In
fact the observed H${\beta}$ equivalent width values and colors
require a large contribution from previous stellar generations
\citep{ter04,mmm08a}.  The evolution of EW(H${\beta}$) {\sl vs.} a
pseudo-color of the continuum, the intensities of the adjacent
continua of [OII]$\lambda$3727 and [OIII]$\lambda$5007 lines (similar
to U-V) of our models compared with the data is shown in Figure
\ref{fig7}.  The SAM and HAM cases for $\Delta t = 1.3$\,Gyr are
plotted in the two first panels. It can be seen that to reproduce the
trend of observational data, shifted to red colors at low values of
EW(H${\beta}$) with respect to the SSP predictions, the contribution
to the total continuum of the non-ionizing red population must be much
higher than the contribution of the current burst which creates the
emission line spectrum. This trend can not be reproduced just with
SSPs neither by increasing their metallicity or age separately nor
simultaneously (MMDT). A strong attenuation in the SFR is needed, as
can be seen in the center panel, to reproduce the whole range in
EW(H${\beta}$) and continuum colors simultaneously.

The time between bursts ($\Delta$t) is a parameter which may also have
an effect on the models similar to the attenuation.  The reduction of
the time between bursts offsets the effect of increasing the
attenuation: The underlying population is less evolved and produces
less reddening.  In Figure \ref{fig7}, right panel, HAM with $\rm
\Delta t = 0.1$\,Gyr and $\rm \Delta t=0.05$\,Gyr are plotted.  The
EW(H${\beta}$) decreases rapidly while resulting colors are not
shifted to the red sufficiently to cover the range shown by the
observational sample. The colors of the models with shorter
inter-burst time are more similar to those with soft attenuation
(strong bursts), and require an extra reddening to reproduce the
needed effects of the underlying non ionizing populations. However,
the EW(H${\beta}$) decreases more from burst to burst than in the case
of a soft attenuation, where it maintains a high value at the beginning
of every bursts. In order to reproduce the observed trend, the
inter-burst time must be longer than 100\,Myr.

In brief,  to reproduce the observable characteristics of 
star-forming galaxies we have to adjust the
three input parameters:
\begin {itemize}

\item  The initial efficiency must be
between 0.10 and 0.60 in order to produce a first burst which provides
oxygen abundances within the observed values. In fact, efficiencies lower
than 10\% seems more probable to reproduce most of observations.

\item Most (perhaps all) H{\sc ii} galaxies require the contribution
of previous stellar generations to explain the observed trends shown
by their continuum and emission line properties. The chemical and
spectro-photometrical parameters (equivalent widths and colors)
obtained by our models reproduce the observed relations if the
contribution of the underlying population from previous bursts to the
total continuum is higher than the contribution of the current burst
of star formation which dominates the observed emission line
spectrum. This implies a history of star formation higher in the past
than at present, and, even in that case, attenuated burst along the time. The
fine tuning of observations is obtained by the adjusting of
the attenuation factor, once the initial efficiency is fixed.
(see Mart\'{\i}n-Manj\'{o}n et al. 2012 for details)

\item The inter-burst time must be shorter than 1.3\,Gyr and longer
than 100\,Myr.  With shorter periods than 100\,Myr the underlying
continuum is too blue and its contribution to the color do not
reproduce the trend shown by H{\sc ii} galaxies.

\end{itemize}

\section{Discussion}

\subsection{The different star formation scenarios.}

Three star formation scenario are usually discussed in relation with
the evolution of dwarf galaxies, (a) Burst: short star-formation
episodes with large quiescent periods, (b) Gasp: long moderate
star-formation episodes with short quiescent periods and (c) Continuous
or almost continuous star formation with few over-imposed sporadic
bursts. The three cases can be simulated by our models simply with
a change of parameters:

\begin{itemize}

\item The star-bursting scenario (a) is this one shown in the previous
sections.  The effects of an instantaneous burst can be seen in the
emission lines during 10\,Myr after the burst takes place. Long
quiescent periods or, in our case, null star formation periods, have
been considered to last more than 1\,Gyr, which would be the minimum
age of the underlying population belonging to the previous
burst. Bursts occurred before the immediately previous one have very
poor contribution to the current total continuum luminosity, and the
use of this long time for the inter-burst period produces a similar
result to assuming a two-burst model for some parameters, as
EW(H${\beta}$). However, differences appear between both scenarios
since the ISM is enriched by every single burst occurring during the
galaxy life time, stellar populations older than 1\,Gyr should be
present, and in consequence colors would be shifted to the red.

\item The gasping star formation scenario (b) can be simulated by
substantially increasing the attenuation of the burst and reducing the
inter-burst time. With the change in these two parameters we make a smoother and more moderate the
star formation. Since the subsequent bursts are
closer in time, the quiescent periods are shorter. Models with
extreme attenuation and inter-burst times shorter than 100\,Myr can be
considered a good approximation to gasping models.  This particular
choice of parameters has important consequences when compared with the
observations. Increasing the attenuation a major contribution to the
continuum by the previous bursts is obtained. However, if, in
addition, the inter-burst time is reduced, the underlying population
will become younger and bluer. This approach would predict the
presence of intermediate age stellar populations, younger than 1\,Gyr,
even after several star bursts. With larger inter-burst times only the
immediately previous stellar generation contributes substantially to
the present continuum, but even older stellar generations would be
noticeable if we further reduce the time between bursts. Although
complicated, it is also possible to simulate more extended star-burst
phases by concatenation of several star-bursts. However, population
synthesis modeling of BCD spectra does not favor extended periods of
star formation \citep{mas99}.  Then, if the starburst phase is short,
there should be many of these episodes during the galaxy lifetime.
According to \cite{salmeida08}, there should be one BCD phase each 0.3\,Gyr, 
which agrees with our models.

\item Under a continuous star formation scenario (c), the SFR must be
very low and extended. If we reduce the inter-burst time to a minimum
value and, simultaneously, we reduce the intensity of each burst, we
obtain a very low star formation rate, not comparable to a starburst, but
similar to a continuous star formation history. 
In fact, many of the observable features of star-forming galaxies can 
be modeled  with a young stellar population 3-5\,Myr old and most
massive stars being essentially coeval. However, as \citet{mas99}
demonstrated, it is also possible to obtain similar results with a
continuous star formation, lasting at least 20\,Myr since the beginning
of the star formation. A low continuous star formation rate cannot be
neglected, especially in low metallicity galaxies. Otherwise, if the SFR
is very low, \citep[log (SFR) $<$ -3 approximately, according
to][models]{mmm09phd}, there are not enough ionizing photons to
produce emission lines. In this case we are not reproducing the
so-called H{\sc ii}  or BCD galaxies, but another type of galaxy or another
stage of their evolution.

\end{itemize}

\subsection{Connecting star-bursting models with the different evolutionary phases of a star-forming dwarf galaxy. }

Actually, we may consider that, instead of a different scenario, the low and
continuous star formation could be the phase occurring between bursts. In
this way we could change the perspective of the problem and study the
characteristics of the galaxies for the two phases: the burst and the
following $\sim 10$\,Myr, and the inter-burst, low activity periods,
that is, after the first 10\,Myr from the last star formation episode.
In fact this is related with the two different time scales involved in
the evolution, and related with different kind of data: On the one hand, a
short time-scale in which the emission lines  and othe effects of the ionizing stellar populations are observed. On the other hand
there is long time-scale which defines the red color evolution mainly due to the
the stellar populations older than 10\,Myr.

Since a H{\sc ii} galaxy can be considered just a
star-forming phase or a stage in the evolution of a gas rich dwarf
galaxy, we can consider the inter-burst time in our
models as a quiescent star-forming dwarf (QBCD) phase
\citep{salmeida08}.  Then the H{\sc ii} galaxy  may be in the phase
10$^{7}$ yr after the star burst, in which its effects are still
visible (emission lines or changes in metallicity due to massive stars
ejections) and the QBCD phase would correspond to the inter-burst
periods. The contribution of the young and old stellar populations in
each galaxy may be obtained from the comparison of data coming from
the two different time-scales, in particular, from observations as
emission lines, proceeding from the ionizing stellar populations, with
colors or abundances given by the older than 10\,Myr stars.

As we have already seen in our previous works, MMDT and \cite{mmm08b}, and we
also show in Fig.~\ref{fig6}, when both types
of stellar populations are contributing to the light, as occurs in BCD 
galaxies, the colors are contaminated by the emission lines,
showing in most cases different trends than those standard sequences
defined by the stellar populations, either young or old. Taking into account
this contamination, the colors observed are those corresponding to the
ionizing continuum of a given galaxy, always during the BCD phase,
that is, the colors of the young population which dominates the
emitted light. In order to study the colors in the inter-burst
periods and to know the properties of the underlying population, we
should further eliminate the continuum arising from the starburst and
isolate the properties of the underlying host galaxy.  

\begin{figure*}
\resizebox{\hsize}{!}{\includegraphics[angle=0]{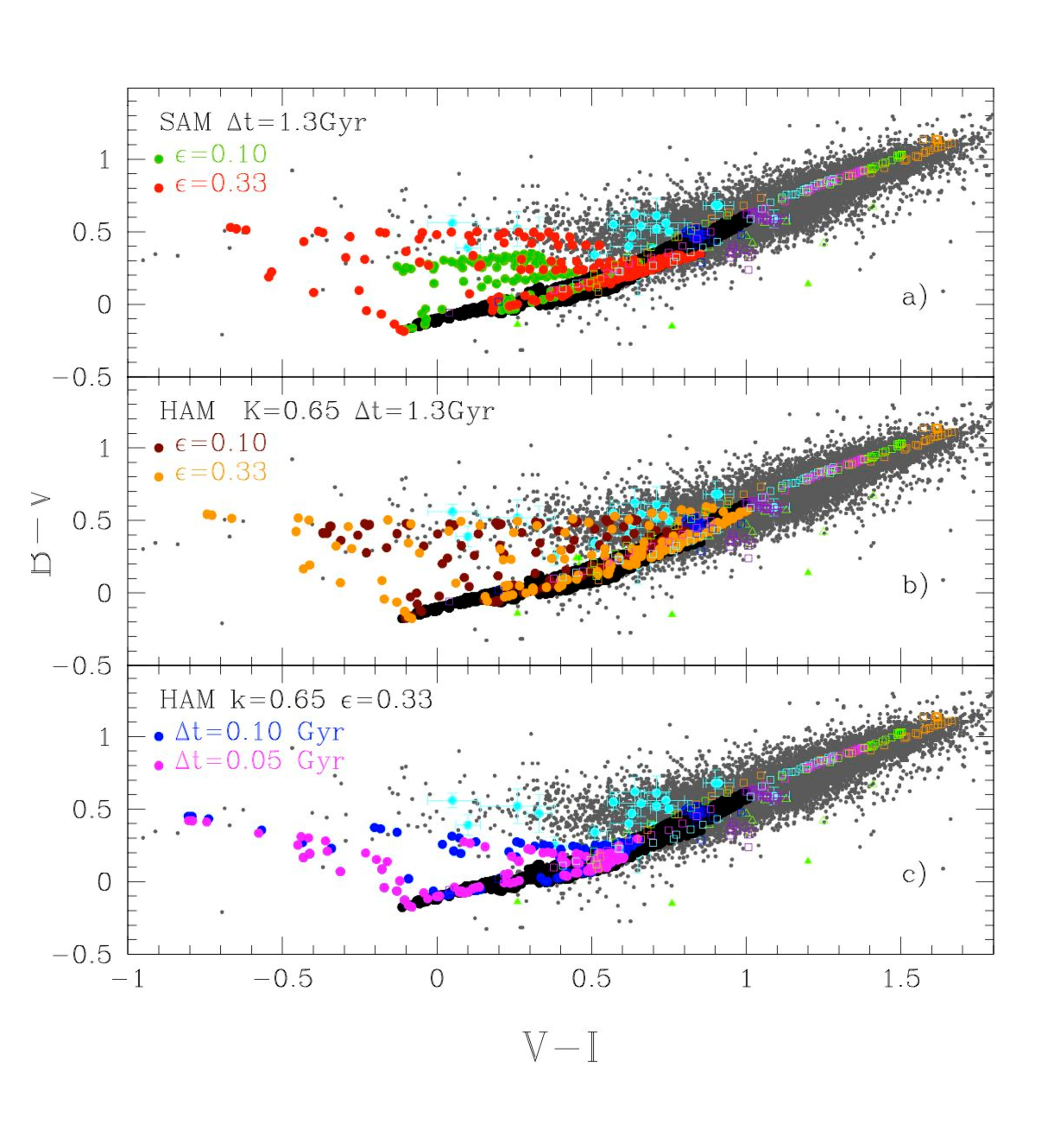}}
\caption{B-V {\sl vs.} V-I color-color diagram for our models including
the stellar continuum colors of the starburst phase (black dots), and
these continuum colors contaminated by the emission lines. In the top panel
we have green and red dots for low and high efficiency SAM models. In the middle panel,
models 3 and 4 as brown and orange dots; And in bottom panel models 5
and 6 with blue and magenta dots. Colors of the inter-burst periods (age $>$ 10 \,Myr) for 6
metallicities are the open (colored) squares. Observational data are taken
from \citet{salmeida08} -grey small dots- and from \citet{cai02,cai011} -cyan full dots
and green open and fill triangles (see text for an explanation).}
\label{fig8} 
\end{figure*}

\begin{figure*}
\resizebox{\hsize}{!}{\includegraphics[angle=0]{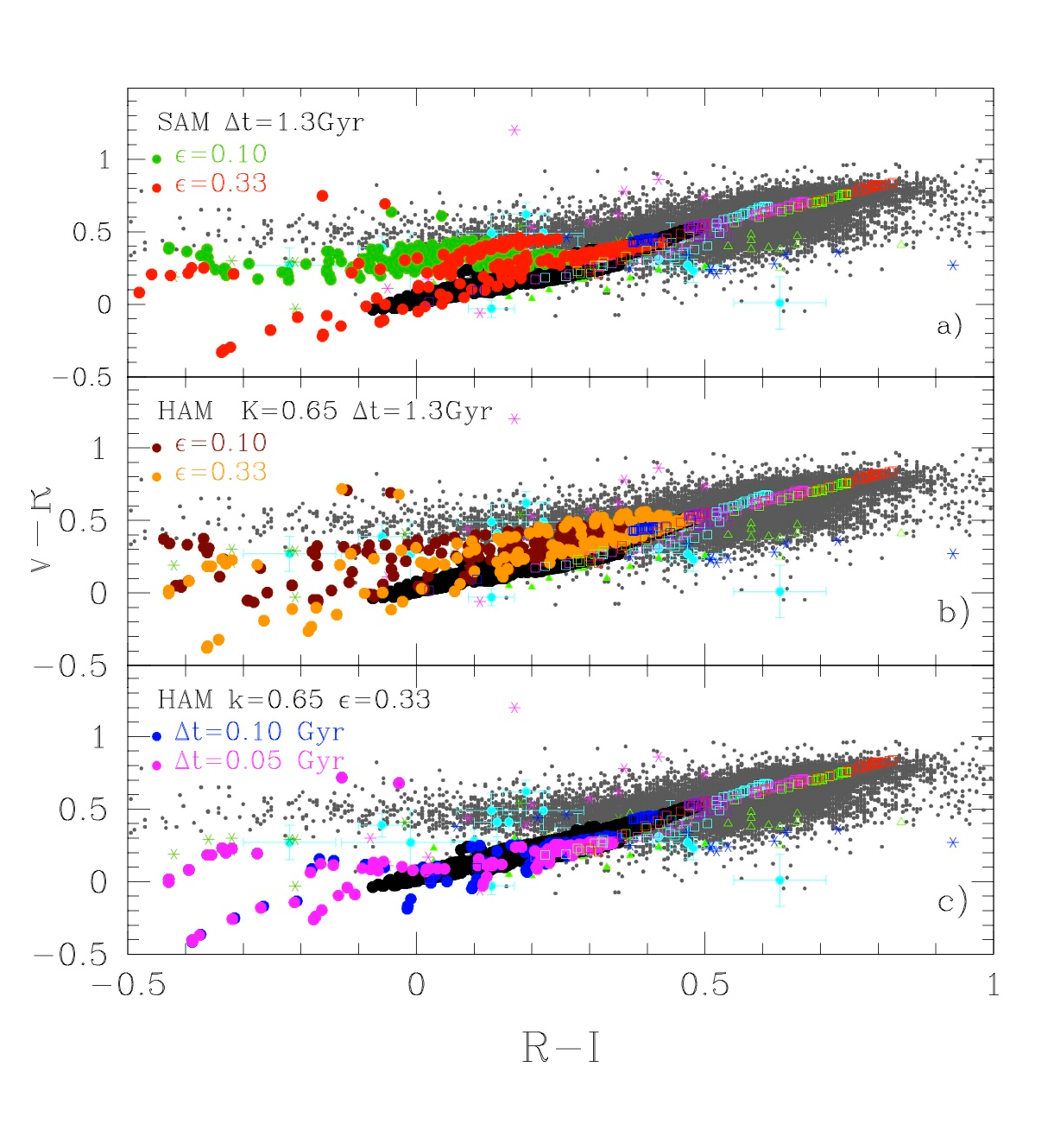}}
\caption{V-I {\sl vs.} V-R color-color diagram for our models
including the stellar continuum colors (black dots), and the continuum
colors contaminated by the emission lines.  In the top panel
we have green and red dots for low and high efficiency SAM models. In the middle panel,
models 3 and 4 as brown and orange dots; And in bottom panel models 5
and 6 with blue and magenta dots. Colors of the inter-burst periods (age $>$ 10 \,Myr) for 6
metallicities are the open (colored) squares. Observational data are taken
from \citet{salmeida08} -grey small dots- and from \citet{cai02,cai011} -cyan full dots
and green open and fill triangles , and   \citet{tell97} --green and magenta stars-- ( see text for explanations).}
\label{fig9}
\end{figure*}

\begin{figure*}
\resizebox{\hsize}{!}{\includegraphics[angle=0]{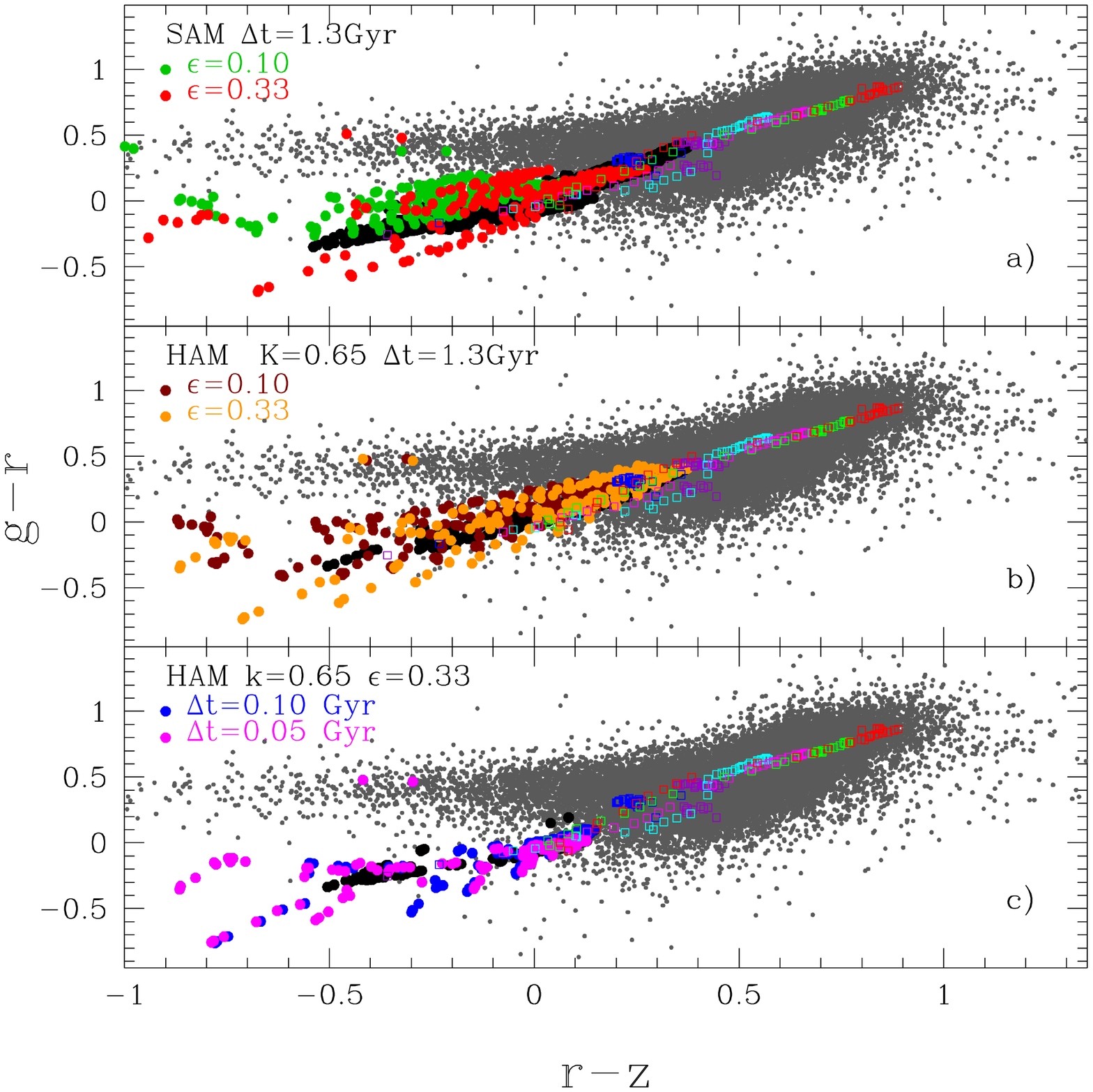}}
\caption{g-r {\sl vs.} r-i color-color diagram for our models including
the stellar continuum colors (black dots), and the continuum colors
contaminated by the emission lines. In the top panel
we have green and red dots for low and high efficiency SAM models. In the middle panel,
models 3 and 4 as brown and orange dots; And in bottom panel models 5
and 6 with blue and magenta dots. Colors of the inter-burst periods (age $>$ 10 \,Myr) for 6
metallicities are the open (colored) squares. Observational data are taken
from \citet{salmeida08} as small grey dots}
\label{fig10} 
\end{figure*}

Figures \ref{fig8}, \ref{fig9} and \ref{fig10} show the B-V {\sl vs.}
V-I, V-I {\sl vs.} R-I and g-r {\sl vs.} r-z color-color diagrams for
our 6 models.  Black circles show the stellar continuum colors in all panels, while the colors computed including the contribution by the strongest emission lines are shown as
green and red for models 1 and 2 respectively, brown and orange for models 3 and 4, and blue and magenta for models 5 and 6.  In order to represent these older phases we
have also represented, as colored open squares, the colors corresponding to
SSPs, as given in MGVB09, for ages older than 10\,Myr that is, without
taking into account the youngest stellar populations phases, and metallicities from Z=0.0001 to 0.02.  We have
not computed exactly the colors of our inter-burst periods, but they
must fall in this same region of the diagrams.

In the B-V {\sl vs.} V-I diagram (Fig. ~\ref{fig8}) the inclusion of
the emission lines contribution to the continuum colors shifts the
position of the model points almost perpendicularly to the originally
computed ones. The location of the points is mainly determined by the
contribution of strong [OIII]$\lambda\lambda$5007,4959 emission lines
to the \textit{V} band, and, logically, they are closer to the normal
stellar populations locus when efficiency is low.  This means that the
excursion from locus defined by the main sequence for stellar populations
is stronger when the star formation efficiency is higher, leading the
points far.  We also show in this figure the
data by \citet{cai011,cai012} as cyan dots with error bars.  This set
of observations corresponds to readily observed colors, including both
continuum and line emission, and no reddening correction has been
applied.  It can be seen that some of the data are impossible to be
reproduced by the models which do not take into account the
contribution by emission lines, even if some amount of reddening is
invoked. On the other hand the data by
\citet{cai02} of resolved locations in Mrk~370, shown in the figure as
open green triangles are very well reproduced by our continuum colors.  The
data of Mrk~370 are based on \textit{UBVRI} broadband and H${\alpha}$
narrow band observations. In this case, the authors subtracted the
contribution of the underlying continuum from the old stellar
population, removing the contribution from emission lines and
correcting for extinction, measuring in this way true colors of the
young star-forming knots. This way while open green triangles correspond to the
observed Mrk~370 colors, including the emission lines contribution and
the underlying stellar population, the full green triangles are the stellar population
colors after correction of extinction, emission lines and old stellar populations. 
In this case we see that some points fall out of the expected region of models
which we interpret as probably due to a over-correction with SSPs models.
The underlying component has colors redder than the BCD ones, and
they are also well reproduced by our models.

In Fig.~\ref{fig9} we plot V-I {\sl vs.} V-R, with the same data  and code from
\citet{cai011,cai012,cai02} and \cite{salmeida08} than in the previous
figure. Moreover, we have also included the data of BCDs and its
hosts from \citet{tell97}. This sample consists of 15 BCDs for which
they observed the total colors. The data corrected of extinction are the 
green stars. When they correct of the emission lines effect, the obtain
the colors shown by magenta stars. These points corresponding to
the actual young stellar populations are  closer to the region where our models are
represented by the same coloured dots as in Fig.~\ref{fig8}. 
They subtracted the contribution of the underlying galaxy by
assuming the mean surface brightness of the extensions, or hosts, to
be constant and representative of the underlying galaxy within the
starburst regions. Summarizing, they give separately the colors of this
underlying stellar continuum (blue stars) within the starburst, the
contributions including the emission lines (green stars) and the colors of
the BCD stellar continuum (magenta stars), corrected from emission lines,
which are close to black dots, corresponding to our continuum BCD
phase. The colors of the isolated host galaxy (blue stars) show redder
colors than the BCD component, in agreement with QBCDs, Mrk~370 host
and model colors with ages older than 10\,Myr.
Again we note some differences with our models that we assign to
an over-correction of the emission lines.

Finally we represent in Fig.\ref{fig10} the g-r vs. r-z diagram with
the same symbols that before. 
The selected QBCDs \citep{salmeida08} are assumed to be like BCD host galaxies \citep{amor07, amor09}. The colors of the SSPs
with ages older than 10\,Myr (colored squares) can reproduce these observations,
indicating that they correspond to the colors of a more evolved
population than the one corresponding to the current burst in
BCDs. However, some of the observed dots lie in the BCD zone,
indicating that this data sample also contains a certain proportion of
BCDs, following a continuous evolutionary sequence.
In this color-color diagram there is a larger difference  than in the previous
ones between models and data for the bluest points, showing these data 
higher g-r values than predicted for similar low r-z colors. 
Probably this is due to the transformation used to calculate
SDSS colors with our Johnson colors,  which would be not valid for this type
of objects with emission lines contaminating the wide band filter magnitudes.

We see in the three figures that models start to separate off the main
sequence of colors at different points following the type of
attenuation. In the top panel, there are points with large
contribution of lines starting even at the lowest and left corner,
where the youngest stellar populations lie. The same points in medium panel
begin to separate at redder colors, this implies that the underlying continuum
corresponds to an older age. These models cover all observational range.
However in the bottom panel the points with the contribution of the lines 
fall in a region much bluer where there are no observations. The underlying
stellar populations are much younger than the populations of the data.

In any case it seems that BCDs and QBCDs colors are well reproduced by
the same basic models, during the first 10\,Myr after the beginning of star
formation and after 10\,Myr up to more than 1\,Gyr respectively, and
therefore we can say that both types of galaxy colors overlap and could be
considered as different phases in the evolution of the same object.

\section{Summary and Conclusions}

Historically, the most common scenario assumed for the star formation
history of the star-forming galaxies is that stars form in a bursting
mode. This scenario consists in successive instantaneous star
formation bursts interposed between long periods of null or very low
star formation activity. If these periods are shorter and the bursts
are more extended and moderate, we are under a gasping star formation
scenario. On the contrary, if the star formation is low and moderate
during the whole life of the galaxy, and only some sporadic bursts
take place, a continuous star formation scenario is appealed.

We have analyzed the possibilities of these scenarios by means of
theoretical models based on the successive bursts star formation
hypothesis. Each galaxy is modeled assuming an initial amount of
unprocessed gas of 10$^{8}$ M$_{\odot}$. The evolution is computed
along a total duration of 13.2\,Gyr during which successive star-bursts
take place.  Since any of the possible scenarios of the star formation
histories described show evidences of being able to explain the
observed data, we have tried to check the hypotheses suggested.

The grid of theoretical models are computed by the combination of
three tools: a chemical evolution code, an evolutionary synthesis code
and a photo-ionization code, all of them previously calibrated. They
have been used in a self-consistent way, {\it i. e.}  taking the same
assumptions about stellar evolution and nucleosynthesis, and the
resulting metallicity in every time step.  These models have three
free input parameters which can be changed to obtain different model
results: (a) The initial star formation efficiency which determines
the initial star formation rate (SFR) and the initial metallicity of
the gas, (b) the attenuation of the successive bursts, which
determines the evolution of the gas, keeping metallicity between the
observational data limits, and (c) the inter-burst time, which sets
the age of the non ionizing underlying population.

We have shown that (a) also leads the behavior of the ionizing gas
since emission lines are determined by the number of massive stars and
the metallicity of the emitting gas. With this parameter we can
control the number of old pristine stars that we are going to find in
the next star formation burst.

The strength of the bursts is controlled by (b). The more attenuated,
the less the star formation rate, which favors the previous stellar
generations to have more weight on the total current spectrum, hence
modifying the equivalent width and the colors of the continuum.

Finally, the contribution of the underlying population can be
determined by (c). These periods are the quiescent phases of the BCD. 
The shorter the inter-burst time, the younger the underlying stellar
populations. These stars will be part of the host BCD galaxy, that is,
all those generations of stars which have not ionizing stars, showing
ages of more than 10$^{7}$ yr.

The models can reproduce simultaneously the data relative to the
current ionizing population and the data which give us information
about the evolutionary history of the star formation.  In order to
reproduce the observational data, the parameters are constrained:
\begin{itemize}
\item[a)] the initial star formation efficiency must be lower than 50\% and more probably
is around 20\% .
\item[b)]  the subsequent star-forming burst must be attenuated, in order to have the adequate proportion of underlying to
young/ionizing stellar populations; and
\item[c)] the inter-burst time must
be longer than 100\,Myr but shorter than 1.3\,Gyr.
\end{itemize}

With our models we are able to compute the stellar continuum colors as
well as the broad band colors corrected from the contribution of the
more intense emission lines. This effect shifts the colors almost
perpendicularly to the stellar continuum ones, specially during the
first bursts of star formation, when the emission lines are more
intense. The contamination of the continuum by the emission lines can
make the observed galaxies appear younger than they really are, hiding
evidence of underlying old populations.

On the other hand, the different characteristics shown by the star
formation histories of our models, imply that we may be observing the
same objects in different evolutionary stages, that is, just during
the burst, in the post-burst phase or in the inter-burst periods.  If
we consider that the different star formation scenarios correspond to
different phases in the life of a single galaxy, our models are able
to reproduce the characteristics of each one of these phases.
Considering the BCD phase as the first 10\,Myr after a star formation
burst, and the quiescent phase as the inter-burst periods, we can
reproduce the galaxies both in a high activity and in a post-starburst
(or pre-starburst) phase. The main difference between these phases are
the existence of emission lines produced by the ionized gas. Their
intensity, and their contribution to the total continuum color, can be
modified by the initial efficiency. With the attenuation factor we can
produce a more or less reddened stellar continuum due to the
contribution of the host galaxy, the age of these old stars being
determined by the inter-burst time.

\section{Acknowledgments}

This work has been partially supported by DGICYT grants
AYA2007--67965-C03-03 and AYA2010--21887--C04--02.  Also, by the Comunidad de
Madrid under grant S-0505/ESP/000237 (ASTROCAM) and by the Spanish MICINN under the Consolider-Ingenio 2010 Program grant CSD2006-00070: First Science with the GTC  (http://www.iac.es/consolider-ingenio-gtc) which are acknowledged. 
RT is grateful to the Mexican Research Council (CONACYT) for supporting 
this research under grants CB-2006-49847,  CB-2007-01-84746 and 
CB-2008-103365-F

\end{document}